\documentclass[runningheads]{llncs}

\usepackage[T1]{fontenc}

\usepackage[table,xcdraw]{xcolor}
\usepackage{graphicx}
\usepackage{makecell}
\usepackage{amsfonts}
\usepackage{tcolorbox}
\usepackage{diagbox}
\usepackage{algorithm}
\usepackage{graphicx}
\usepackage{tabularx}
\usepackage{enumitem}
\usepackage{booktabs}
\usepackage[frozencache=true,cachedir=minted]{minted}
\usepackage{caption}
\usepackage{mathtools}

\usepackage{pifont}
\usepackage{hyperref}

\newfloat{steps}{htbp}{lop}
\floatname{steps}{STEPS}

\begin{document}

\title{CASTLE: Benchmarking Dataset for Static Code Analyzers and LLMs towards CWE Detection}
\titlerunning{CASTE benchmark}

\author{Richard A. Dubniczky\inst{1} \and Krisztofer Zoltan Horvát\inst{1} \and Tamás Bisztray\inst{2,3} \and Mohamed Amine Ferrag\inst{4} \and Lucas C. Cordeiro\inst{5,6} \and Norbert Tihanyi\inst{1,7}}
\authorrunning{R. A. Dubniczky, et.al.}
\institute{%
    \inst{} Eötvös Loránd University (ELTE), Budapest, Hungary \\
    \and
    \inst{} University of Oslo, Oslo, Norway \\
    \and
    \inst{} Cyentific AS, Oslo, Norway \\
    \and
     \inst{} Guelma University, Guelma, Algeria \\
    \and
    \inst{} The University of Manchester, Manchester, UK \\
    \and
    \inst{} Federal University of Amazonas, Manaus, Brazil \\
    \and
    \inst{} Technology Innovation Institute (TII), Abu Dhabi, UAE \\
}


\maketitle

\begin{abstract}

Identifying vulnerabilities in source code is crucial, especially in critical software components. Existing methods such as static analysis, dynamic analysis, formal verification, and recently Large Language Models are widely used to detect security flaws. This paper introduces CASTLE (CWE Automated Security Testing and Low-Level Evaluation), a benchmarking framework for evaluating the vulnerability detection capabilities of different methods. We assess 13 static analysis tools, 10 LLMs, and 2 formal verification tools using a hand-crafted dataset of 250 micro-benchmark programs covering 25 common CWEs. We propose the CASTLE Score, a novel evaluation metric for fair comparison. Our results reveal key differences: ESBMC (a formal verification tool) minimizes false positives but struggles with vulnerabilities beyond model checking, such as weak cryptography or SQL injection. Static analyzers suffer from high false positives, increasing manual validation efforts for developers. LLMs perform exceptionally well in the CASTLE dataset when identifying vulnerabilities in small code snippets. However, their accuracy declines, and hallucinations increase as the code size grows. These results suggest that LLMs could play a pivotal role in future security solutions, particularly within code completion frameworks, where they can provide real-time guidance to prevent vulnerabilities. The dataset is accessible at \url{https://github.com/CASTLE-Benchmark} .


\end{abstract}

\section{Introduction}

Rapid advancements in artificial intelligence (AI) have sparked both excitement and concern about the future of traditional software engineering. For instance, Meta’s recent announcement that AI could soon replace many software engineering roles highlights a shifting landscape in code development~\cite{marks2025zuckerberg}. While AI-driven code generation offers remarkable efficiency, a study by Tihanyi et al. found that all examined large language models (LLMs) produced vulnerable C code~\cite{tihanyi2024secure}. Similar conclusions have been reached in studies examining other programming languages, such as PHP and Python~\cite{10.1007/978-3-031-68738-9_34,MECHRI2025104151}. These large-scale studies consistently indicate that such vulnerabilities arise partly because LLMs lack contextual understanding during the generation process. 
Several studies highlight that once the code is generated, and a vulnerability is identified, LLMs excel in fixing these~\cite{InterFix,tihanyi2024newerasoftwaresecurity}. The question is, how do we find the vulnerabilities? Despite the growing importance of automated software verification, developers and security practitioners lack clear guidance on which tools are most reliable for detecting vulnerabilities in C code. Several interrelated issues contribute to this uncertainty:

\begin{enumerate}[noitemsep,topsep=3pt] \footnotesize 
\item \textbf{Diverse vulnerability types.} Security flaws in C code range from classic memory management issues (e.g., buffer overflows) to subtler logical errors. We need to understand which detection methods can reliably detect different categories.
\item \textbf{Emergence of LLMs.} While LLMs exhibit promise in automated code generation, bug fixing, and vulnerability detection, their reliability in different vulnerabilities and coding scenarios is unclear. 
\item \textbf{Lack of standardized benchmarks.} Existing datasets often contain too many samples with imbalanced CWE representations, and fail to represent the breadth of CWE vulnerabilities. Tools that rely on compilable code—particularly FV methods—are especially disadvantaged without realistic, fully functional programs. To gauge each tool’s performance accurately, a benchmark must be rigorously validated, contain clearly labeled vulnerabilities, and support line-level detection granularity. 
\end{enumerate}

Given these challenges, a robust and compilable benchmark dataset that accurately captures major CWE vulnerabilities is paramount to answer the following research questions: 
\begin{tcolorbox}[colback=gray!10]

\begin{itemize}
\item {\textbf{RQ1}:} How do state-of-the-art static analysis tools, formal verification methods, and LLM-based approaches compare to effectively detecting C code vulnerabilities?
\item {\textbf{RQ2}:} Are combinations of tools more effective than using a single 
tool?
\item {\textbf{RQ3}:} What metrics can reliably demonstrate these differences among various tools?
    
\end{itemize}

\end{tcolorbox}
Our study holds the following contributions:
\begin{itemize}[topsep=3pt] \footnotesize 
\item  We introduce \textbf{CASTLE (CWE Automated Security Testing and Low-Level Evaluation)}, a curated collection of $250$ compilable, compact C programs, each containing a single CWE. This benchmark is aimed at enabling direct comparisons among current and future vulnerability scanning tools, including traditional static analyzers, FV techniques, and LLM-based approaches. 
\item We conduct a broad comparison of the most widely used static code analyzers and popular LLMs to assess their effectiveness in detecting important vulnerabilities in the C language, using a new metric called  \textbf{CASTLE Score}, thereby providing crucial insights into their relative strengths and weaknesses. \end{itemize}

The rest of this work is structured as follows: Section~\ref{sec:related} reviews related literature and outlines the current state of vulnerability scanning tools and AI-based code analysis. Section~\ref{sec:methodology} details the construction of the CASTLE benchmark, including the selection criteria for CWEs and the methodology for creating the curated C programs. Section~\ref{sec:discussion} discusses the results, and presents the experimental setup and comparative analysis of the 11 static code analyzers and 11 LLMs. Section~\ref{sec:limitations} overviews limitations. Finally, Section~\ref{sec:conclusion} concludes the paper and outlines potential directions for further research.

\section{Related work}
\label{sec:related}

Ensuring software correctness, safety, and security is central to software engineering. Examining related literature on the role of AI in software development, most of the existing work and benchmarking approaches focused on testing LLMs' capabilities in producing functionally correct code. However, safety and security are just as important. 

\subsection{Datasets and Benchmarks}

Existing vulnerability datasets are frequently used for fine-tuning machine learning models, yet they exhibit several shortcomings that make them unsuitable for comprehensive benchmarking. First, many datasets offer imbalanced representations of CWE categories, failing to provide adequate test coverage of certain vulnerability types. Second, an extreme or uneven distribution of vulnerable versus non-vulnerable samples either hinders accurate false-positive evaluation (when nearly all samples are vulnerable) or fails to capture diverse false-negative scenarios (when some vulnerability types remain underrepresented).

\begin{table}[htb]
\scriptsize
\centering
\addtolength{\tabcolsep}{-2.3pt}
\caption{C/C++ Datasets for Vulnerability Detection}
\begin{tabular}{%
  >{\ttfamily}p{2.6cm}%
  >{\ttfamily}p{1cm}%
  >{\ttfamily}p{1.6cm}%
  >{\ttfamily}p{1.5cm}%
  >{\ttfamily}p{1.5cm}%
  >{\ttfamily}p{1.6cm}%
  >{\ttfamily}p{1.6cm}%
  >{\ttfamily}p{1.3cm}}
\toprule
\textbf{Dataset} & 
\textbf{Size} &
\textbf{\#Multiple Vuln./File} & 
\textbf{Vuln. Snippets } & 
\textbf{Compilable} &
\textbf{Granularity} &
\textbf{Labelling} &
\textbf{Source} \\
\midrule

Draper \cite{russell2018automated} & 1274k & \ding{52}  & 5.62\%   & \ding{56} & function & Stat  & mixed      \\
Big-Vul \cite{fan2020bigvul}       & 264k  & \ding{56}  & 100\%    & \ding{56} & function & Patch  & real-world \\
DiverseVul \cite{chen2023diversevul}
                                   & 349k  & \ding{56} & 7.02\%   & \ding{56} & function & Patch  & real-world \\
FormAI-v2 \cite{tihanyi2024secure}  & 331k  & \ding{52}  & 62.07\%  & \ding{52} & file     & FV & AI Gen. \\
PrimeVul \cite{ding_vulnerability_2024}
                                   & 235k  & \ding{56}& 3\%      & \ding{56}     & function & Manual & real-world \\
SARD \cite{NIST_SARD}              & 101k  & \ding{56} & 100\%    & \ding{52} & file     & B/S/M & mixed  \\
Juliet (C/C++) \cite{2024nistjulietc}
                                   & 64k   & \ding{56} & 100\%    & \ding{52} & file     & BDV  & synthetic \\
Devign \cite{zhou2019devign}     & 28k   & \ding{56} & 46.05\%  & \ding{56} & function & Manual  & real-world \\
REVEAL \cite{9448435}     & 23k   & \ding{56}& 9.85\%   & \ding{56} & function & Patch & real-world \\

CVEfixes \cite{guru2021cvefixes}   & 20k   & \ding{56} & 100\%      & \ding{56}       & commit   & Patch & real-world \\
\bottomrule
\end{tabular}
\label{tab:datasets}
Legend:
\textbf{Patch}: GitHub Commits Patching a Vulnerability, \textbf{Stat}: Static Analyzer, \\
\textbf{BDV}: By Design Vulnerable, \textbf{FV}: Formal Verification with ESBMC, \textbf{Manual}: Manual Labeling by Human Experts
\end{table}

Furthermore, a key challenge is that many popular datasets lack compilable programs, making it impossible to meaningfully assess formal verification tools such as the \textit{Efficient SMT-based Context-Bounded Model Checker (ESBMC}). Among compilable options, SARD stands out for including the Juliet test cases and $45,437$ C samples mapped to CWE categories. However, some files exceed $3,000$ lines of code, creating two constraints: \begin{enumerate}[noitemsep,topsep=3pt] \item Large token sizes impose high computational costs on LLM-based approaches and limit the use of smaller-parameter models. \item The complexity and volume of large files can overwhelm formal verification tools, dramatically increasing runtime and impeding direct comparisons with other analyzers. \end{enumerate} \noindent Another example is FormAI, a large-scale dataset labeled using ESBMC itself. As a result, it excludes crucial vulnerability classes, such as cross-site scripting (XSS), SQL injection or OS command injection,  which exceed the capabilities of current FV tools.

In contrast, CASTLE provides a collection of compilable code snippets, deliberately crafted to cover major CWEs while minimizing the number of queries required for effective analysis. This design enables the straightforward deployment of LLM-based methods and traditional static analyzers with specialized wrappers, facilitating rapid, automated evaluation across various tools. Additionally, the newly introduced \textit{CASTLE score} provides a more detailed comparative metric than conventional pass/fail assessments, allowing for clearer differentiation of subtle performance variations among state-of-the-art tools. The CASTLE dataset balances vulnerable and non-vulnerable samples, permitting more robust evaluations of false positives and negatives.

\subsection{Traditional Vulnerability Scanning Overview}

Traditional approaches have long relied on static analysis methods, such as pattern matching, data flow analysis, and taint analysis, as well as dynamic analysis techniques like fuzz testing~\cite{fuzzing}. Likewise, \textit{Formal Verification (FV)} methods~\cite{Formalverification}, including \textit{Bounded Model Checking (BMC)}~\cite{BMC} and theorem proving, are widely employed to detect security flaws such as buffer overflows. The NIST-led Static Analysis Tool Exposition (SATE)~\cite{okun2009static,nist-sate-vi} provided large-scale evaluations on open-source code, confirming that while these scanners could spot certain weaknesses, no single method excelled across all vulnerability types.

Academic and industrial benchmarks reveal similar shortcomings. Early work by Wilander and Kamkar~\cite{wilander2002comparison} showed that five tools missed most \texttt{C} function vulnerabilities and produced many false positives, a trend later echoed by Emanuelsson and Nilsson~\cite{emanuelsson2008comparative}. Johns and Jodeit~\cite{johns2011scanstud} demonstrated synthetic benchmarks to distinguish genuine alerts from false alarms, while Bennett~\cite{bennett2024semgrep} reported detection rates of 11.2\%--26.5\% for standard SAST tools, improved to 44.7\% by augmenting them with enhanced Semgrep rules.

\subsection{LLM-Based Vulnerability Detection}

Recent years have witnessed a growing interest in using Large Language Models (LLMs) for automated vulnerability detection~\cite{YANG2025112234,li2024llm,lee2024improving}. Although these models are often praised for handling diverse code repositories, they primarily rely on pattern-based sequence learning rather than (neuro-)symbolic reasoning. As a result, LLMs can detect certain coding flaws effectively, yet they remain susceptible to overlooking complex or context-dependent vulnerabilities.

Recent developments, particularly in decoder-only models such as OpenAI’s ChatGPT and Meta’s Code Llama, highlight a shift in how researchers and practitioners approach vulnerability detection. Their larger context window and on-demand text generation facilitate powerful few-shot or prompt-based strategies that, for specific benchmarks, surpass classical fine-tuned detectors. For instance, properly designed chain-of-thought prompts have been reported to increase F1 scores on real-world vulnerabilities by providing step-by-step guidance for analyzing the code~\cite{MECHRI2025104151}.

Vulnerability detectors typically leverage transformer-based code models trained on massive code corpora spanning multiple languages. These training datasets often contain their share of insecure code, potentially introducing biases or ``model collapse'' issues~\cite{shumailov}. Broadly, transformer models are categorized into three groups~\cite{sheng2025large}:
\begin{enumerate}
    \item \textbf{Encoder-Only}: Used for classification tasks. Early work on vulnerability detection often fine-tuned these models to label code snippets as “vulnerable” or “safe.” They generally require full retraining for each new task.
    \item \textbf{Encoder–Decoder}: Useful for sequence-to-sequence tasks, such as code summarization or refactoring, but they can also be adapted for classification. 
    \item \textbf{Decoder-Only}: Increasingly favored due to large context windows and flexible in-context learning. These models can be prompted to identify vulnerabilities (and sometimes even propose potential fixes) without parameter updates, relying on the knowledge captured during pre-training.
\end{enumerate}

The trend toward decoder-only architectures aligns with industry practices, where state-of-the-art LLMs (e.g., GPT-4) are often served via specialized prompts rather than exhaustive retraining. This approach leverages \emph{in-context learning}, enabling the model to understand and analyze security issues on demand. Carefully constructed prompts—such as chain-of-thought instructions—can improve detection accuracy by guiding the model’s attention toward specific code patterns or CWE categories~\cite{MECHRI2025104151}.

Existing work indicates that LLM-based solutions can outperform traditional static analyzers on well-defined benchmarks~\cite{li2024llm,lee2024improving,YANG2025112234}. However, these improvements do not translate uniformly across all vulnerability types, and use cases: LLMs often fail at detecting nuanced, multi-function flaws or to interpret extensive code segments. 

\section{Methodology}
\label{sec:methodology}

\begin{figure*}[ht] 
\centering
\includegraphics[width=0.9\textwidth]{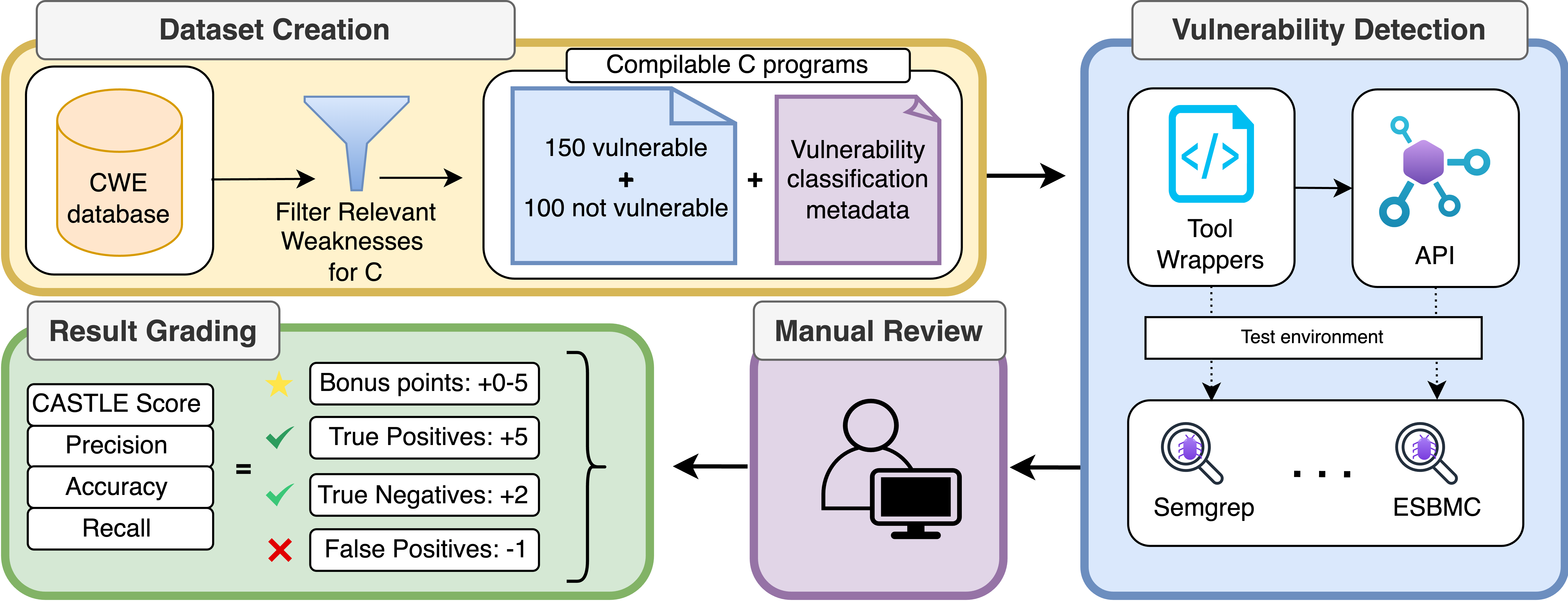}
\caption{The CASTLE Benchmark Framework.}
\label{fig:framework}
\end{figure*}

This section overviews the dataset creation process and introduces our research's newly developed evaluation metrics. Figure~\ref{fig:framework} provides a visual overview of the dataset creation and testing framework. 

\subsection{Dataset}
The CASTLE dataset comprises 250 small programs in C, each crafted manually by cybersecurity experts. It encompasses $25$ distinct CWEs, with $10$ test cases per CWE ($6$ vulnerable and $4$ non-vulnerable). This balanced distribution facilitates focused assessments of each tool’s vulnerability detection capabilities while accurately measuring false positives. In ambiguous cases, experts selected a higher-level CWE category or iteratively refined the test until only the most relevant CWE remained.

Each program was required to compile without errors---although compiler warnings were permitted---as this is a requirement for formal verification tools. All benchmarks were written in C and selected for their capacity to accommodate a wide range of vulnerability types, including intricate memory management issues. Furthermore, each test case was restricted to a single file (with optional external libraries) and designed to contain exactly one or zero vulnerabilities. This structure simplifies the identification of vulnerabilities and helps prevent confusion when validating false positives.

\begin{table}[thb]
    \centering
   \scriptsize
    \begin{tabularx}{\textwidth}{lcX}
        \toprule
        \textbf{CWE} & \makecell{\textbf{Top 25} \\ \textbf{Standing}} & \textbf{Weakness Description} \\
        \hline
CWE-22 & 5 & Improper Limitation of a Pathname to a Restricted Directory  \\
CWE-78 & 7 & Improper Neutralization of Special Elements used in an OS Command  \\
CWE-89 & 3 & Improper Neutralization of Special Elements used in an SQL Command  \\
CWE-125 & 6 & Out-of-bounds Read \\
CWE-134 & 12 & Use of Externally-Controlled Format String \\
CWE-190 & 23 & Integer Overflow or Wraparound \\
CWE-253 & - & Incorrect Check of Function Return Value \\
CWE-327 & - & Use of a Broken or Risky Cryptographic Algorithm \\
CWE-362 & - & Concurrent Execution using Shared Resource with Improper Synchronization  \\
CWE-369 & 23 & Divide By Zero \\
CWE-401 & - & Missing Release of Memory after Effective Lifetime \\
CWE-415 & 21 & Double Free \\
CWE-416 & 8 & Use After Free \\
CWE-476 & 21 & NULL Pointer Dereference \\
CWE-522 & 14 & Insufficiently Protected Credentials \\
CWE-617 & - & Reachable Assertion \\
CWE-628 & - & Function Call with Incorrectly Specified Arguments \\
CWE-674 & 24 & Uncontrolled Recursion \\
CWE-761 & 20 & Free of Pointer not at Start of Buffer \\
CWE-770 & 24 & Allocation of Resources Without Limits or Throttling \\
CWE-787 & 2 & Out-of-bounds Write \\
CWE-798 & 14 & Use of Hard-coded Credentials \\
CWE-822 & 20 & Untrusted Pointer Dereference \\
CWE-835 & 24 & Loop with Unreachable Exit Condition  \\
CWE-843 & - & Access of Resource Using Incompatible Type  \\

        \bottomrule
    \end{tabularx}
        \caption{List of CWEs in the benchmark concerning their presence in MITRE's 2024 Top 25 vulnerability list~\cite{mitre-2024top25}}
    \label{tab:cwe_list}
\end{table}

When incorporating the system prompt alongside the source code, the total input tokens across the dataset amount to approximately 115,620 tokens using the 
\textit{cl100k\_base} encoding scheme. This total reflects the resource considerations required when running evaluations with token-sensitive language models.
The dataset was intentionally capped at 250 benchmarks to make thorough manual verification feasible. This rigorous verification process is indispensable for detecting false positives and confirming line-level detections. Moreover, this selective approach supports the cost-effective evaluation of computationally intensive tools, including advanced LLMs (e.g., GPT-o1, GPT-o3, DeepSeek R1).

The benchmarks exhibit substantial variability in complexity. Code lengths range from $7$ to $164$ lines, yielding $10,392$ lines (an average of $42$ lines). Each includes $1$--$8$ functions ($2.2$ on average), with cyclomatic complexity values spanning 1--29 (mean $6.3$). Halstead volumes range from approximately $89.9$ to over $5,246.7$, averaging $1,104.8$. 

This breadth ensures the dataset covers a wide spectrum of vulnerabilities, from lower-level issues (e.g., memory management flaws, race conditions) to higher-level security risks (e.g., command injections, hard-coded credentials). Most CWEs were chosen based on their prevalence in the Top 25 CWEs of 2023--2024. Each test underwent iterative validation by human experts to ensure overall quality and reliability. Table~\ref{tab:cwe_list} provides a comprehensive list of the included CWEs.

\subsection{Test Format and Wrappers}

Each test in the dataset comprises two components: a metadata block and the source code. Both are stored in a single file for streamlined development and validation, as illustrated in Listing~\ref{lis:test_example}. The metadata, formatted in YAML, precedes the source code and is removed during preprocessing. All lines containing vulnerabilities are marked using the comment string \texttt{// \{!LINE\}}, ensuring consistent identification across different tools. We note that for LLM evaluation, all side-channel information that could introduce bias is removed during the analysis.

Additionally, the metadata specifies the vulnerability's CWE classification and other contextual information. After processing, each test is converted into a JSON-formatted dictionary that includes the code, metadata, and computed software metrics (e.g., cyclomatic complexity, Halstead volume). This unified structure simplifies integration with the various wrappers, facilitating automated execution and standardized result reporting.

\begin{listing}[htbp]
\caption{An example of a micro-benchmark illustrating a buffer overflow}
\label{lis:test_example}
\begin{tcolorbox}[title={CASTLE-787-1.c Test Source Code}, coltitle=white, colbacktitle=black]
\scriptsize
\begin{minted}[xleftmargin=6pt,linenos,escapeinside=||]{c}
/*
================================================
dataset: CASTLE-Benchmark
name: CASTLE-787-1.c
version: 1.1
compile: gcc CASTLE-787-1.c -o CASTLE-787-1
vulnerable: true
description: Buffer overflow in scanf function copying into a fixed length buffer.
cwe: 787
==============================================
*/

#include <stdio.h>

int main(int argc, char *argv[])
{
    char reg_name[12];
    printf("Enter your username:");
    scanf("%s", reg_name); // {!LINE}
    printf("Hello %s.\n", reg_name);
    return 0;
}
\end{minted}
\end{tcolorbox}
\end{listing}

To ensure a uniform and reproducible evaluation across all tools, we developed custom wrappers that automate installation, configuration, execution, and result retrieval.
Each tool was containerized via Docker alongside its dependencies for freely available static analyzers (e.g., Clang Analyzer, ESBMC). We then used Python scripts to run each tool on all test cases, collecting and parsing the results into a standardized report format.

For closed-source solutions such as CodeThreat and Aikido, we uploaded the micro-benchmarks to secure repositories or dashboards accessible via proprietary APIs. The returned results were automatically parsed, and manual consistency checks were performed to verify alignment between reported findings and the tools’ web interfaces.

LLM-based vulnerability detection was driven by a generic script that interacted with standard OpenAI APIs. Each model was prompted to return JSON-formatted detection results. Smaller models (fewer than 6B parameters) often struggled to generate well-structured JSON, suggesting limitations in handling detailed output formats. Additionally, models were sensitive to line-specific detections, occasionally identifying the correct vulnerability but offsetting the line number. We also prompted LLMs to provide the corresponding code lines to address minor positioning errors, allowing minimal adjustments during evaluation.

All wrappers developed for this research are publicly available in the main repository. However, intermediate analysis reports are not provided, as they may include proprietary information protected by the respective tool vendors. Each wrapper saves the results in a custom report format, which is later used to process the results and calculate the metrics for the tools.

\begin{listing}[!htbp]
\caption{Prompt Template for LLMs for JSON-formatted vulnerability reports}
\label{lis:prompt}
\begin{tcolorbox}[title={LLM System Prompt}, coltitle=white, colbacktitle=black]
\scriptsize
\begin{minted}[xleftmargin=6pt,linenos,breaklines,escapeinside=||]{md}
You are a professional security analyst reviewing C code for vulnerabilities.
You will list the found vulnerabilities in a JSON format using the exact template below:

```
[
    {"severity": string, "line": int, "cwe": int, "message": string, "line_content": string},
    ...
]
```
### Rules:
1. Do not omit the triple backticks (``` at the beginning and ``` at the end).
2. If you did not find any vulnerabilities, return an empty list.
3. If you don’t know the CWE number, set it to 0.
4. Any response that does not follow the above format is invalid.
5. You get 5 points for each vulnerability you find, but get -1 for all false positives you report.

Now, review the following C code and return your response:

\end{minted}
\end{tcolorbox}
\vspace{-8pt}
\label{lis:prompt}
\end{listing}

\subsection{CASTLE Score}

In this section, we introduce the \textit{CASTLE} score, a new metric for evaluating the performance of vulnerability detection tools with the CASTLE-Benchmark. The CASTLE score integrates both true- and false-positive rates, awards bonus points for detecting high-impact vulnerabilities (based on the Top 25 CWEs), and rewards correct identification of non-vulnerable code. By incorporating these factors, the metric better captures a tool's overall reliability than standard pass/fail evaluations.

Let $d^n = \{\,d_1, d_2, \dots, d_n\}$ denote a dataset of $n \in \mathbb{N}^{+}$ micro-benchmark tests. Each test $d_i$ targets a specific security weakness (e.g., buffer overflow) or contains no vulnerabilities. Let $v_i$ denote the correct vulnerability label associated with $d_i$. If it does not contain a vulnerability, then $v_i = \emptyset$ For any given tool $t$, let $t(d_i)$ represent the set of vulnerabilities reported by $t$ when analyzing $d_i$.

\textit{Bonus Formula:} Following the Top 25 CWE list released by MITRE \cite{mitre-2024top25}, let $S : \text{CWE} \to \{1, 2, \dots, 25\} \cup \{\infty\}$ be a function that returns the rank of a given weakness if it appears in the top 25 list, with $S(c) = \infty$ assigned to any CWE not in the list. Define $b_{\text{max}} = 5$ as the maximum bonus for detecting a Top-25 CWE. For a found vulnerability labeled $\text{cwe} = t_{cwe}$, the bonus $B(t_{cwe})$ is computed as:
\begin{equation}
    B\bigl(t_{cwe}\bigr) =
        \begin{cases}
        b_{\text{max}} - \left\lfloor \frac{S(t_{cwe}) - 1}{b_{\text{max}}} \right\rfloor, & \text{if } S(t_{cwe}) \leq 25 \\
        0, & \text{otherwise}
        \end{cases}
\end{equation}
Thus, a tool detecting a highly ranked CWE (e.g., Top 5) receives the full bonus of 5 points, while lower-ranked CWEs yield a proportionally reduced bonus. CWEs outside the Top 25 list receive no bonus.

\textit{Scoring Formula}:
For each test $d_i$, a tool's performance is scored according to whether it correctly identifies the vulnerability or the true negative sample. The final CASTLE score for a tool $t$ over the CASTLE benchmark is:
\begin{equation}
    \mathrm{CASTLE}\bigl(t^n\bigr) \;=\;
    \sum_{i=1}^{n}\;
    \begin{cases}
        5 \;-\;\bigl(|t(d_i)| - 1\bigr)\;+\;B\bigl(t_{cwe}\bigr),
            & \text{if } v_i \neq \emptyset \;\land\; v_i \in t(d_i) \\[8pt]
        2, 
            & \text{if } v_i = \emptyset \;\land\; t(d_i) = \emptyset \\[8pt]
        -\,\bigl|t(d_i)\bigr|,
            & \text{otherwise}
    \end{cases}
    \label{eq:castle_score}
\end{equation}

\noindent
\textbf{Interpretation:}
\begin{itemize}[leftmargin=1em, itemsep=2pt, topsep=2pt]
    \item \emph{Correct Vulnerability Detection} If a benchmark is vulnerable ($v_i \neq \emptyset$) and the tool detects exactly that vulnerability, the tool scores 5 points plus an additional bonus $B(t_{cwe})$ depending on the CWE's standing in the top 25. However, multiple reported findings ($|t(d_i)|>1$) reduce the score by one for each, penalizing extraneous detections.
    \item \emph{Correct Non-Vulnerability Detection.} If the benchmark is non-vulnerable ($v_i=\emptyset$) and the tool reports no vulnerabilities, it earns 2 points.
    \item \emph{All Other Cases.} If the tool misses a vulnerability (failing to report $v_i$), or incorrectly flags any vulnerability (including false positives in a non-vulnerable test), the score is reduced by one for each false-positive finding ($-\lvert t(d_i)\rvert$). Notably, it does not incur additional penalties if the tool reports nothing on a vulnerable benchmark.
\end{itemize}

\subsection{CASTLE Combination Score}

An additional feature of the CASTLE score is its applicability to tool combinations. Specifically, if two or more tools exhibit high overlap in detected CWEs, their combined false positives may outweigh any marginal gain from additional true positives, thus lowering the overall score. Conversely, if tools complement each other’s coverage without substantially increasing false-positive rates, their combination can yield higher net performance.

To compute the \textit{CASTLE Combination Score}, one considers the union of reported vulnerabilities and awards true positives and true negatives once while aggregating penalties for all false positives. This ensures that overlapping detections do not artificially inflate the combined score and that the negative impact of extraneous findings remains cumulative.
The combination score can be calculated for an any $n$ number of tool combinations.

\section{Discussion}
\label{sec:discussion}

We evaluated $13$ static code analyzers, $2$ formal verification tools, and $10$ LLMs on the CASTLE benchmark. The results, including the CASTLE Scores, are presented in Table~\ref{tab:results}. Tools and LLMs are distinctly separated, and the reasoning behind this will be discussed in this chapter.

\begin{table*}[htbp]
\scriptsize
\centering
\caption{The results from 250 C tests and their CASTLE Scores}
\begin{tabular}{|ll|cccc|ccc|c|}
\toprule
\multicolumn{2}{c}{} & \multicolumn{4}{c}{\textbf{Results}} & \multicolumn{4}{c}{\textbf{Evaluation Metrics}} \\
\cmidrule(lr){3-6} \cmidrule(lr){7-10}
\textbf{Name} & \textbf{Version}* & \textbf{TP} & \textbf{TN} & \textbf{FP} & \textbf{FN} & \textbf{P} & \textbf{R} & \textbf{A} & \textbf{CASTLE Score} \\
\midrule
ESBMC & 7.8.1 & 53 & 91 & 32 & 97 & 62\% & 35\% & 53\% & 661 \cellcolor[rgb]{0.6941, 0.8696, 0.4436} \\
CodeQL & 2.20.1 & 45 & 79 & 49 & 112 & 48\% & 29\% & 44\% & 634 \cellcolor[rgb]{0.7490, 0.8933, 0.4791} \\
GCC Fanalyzer & 13.3.0 & 41 & 76 & 93 & 109 & 31\% & 27\% & 37\% & 559 \cellcolor[rgb]{0.8919, 0.9545, 0.6011} \\
Snyk & 1.1295.4 & 26 & 82 & 42 & 124 & 38\% & 17\% & 39\% & 552 \cellcolor[rgb]{0.9036, 0.9594, 0.6171} \\
Semgrep Code & 1.110.0 & 36 & 73 & 70 & 120 & 34\% & 23\% & 36\% & 541 \cellcolor[rgb]{0.9211, 0.9668, 0.6411} \\
CBMC & 5.95.1 & 18 & 100 & 0 & 132 & 100\% & 12\% & 47\% & 536 \cellcolor[rgb]{0.9270, 0.9692, 0.6491} \\
SonarQube & 25.3.0 & 43 & 68 & 135 & 107 & 24\% & 29\% & 31\% & 511 \cellcolor[rgb]{0.9679, 0.9865, 0.7050} \\
Aikido & - & 14 & 83 & 40 & 136 & 26\% & 9\% & 36\% & 481 \cellcolor[rgb]{0.9998, 0.9928, 0.7370} \\
Jit & - & 21 & 78 & 68 & 134 & 24\% & 14\% & 33\% & 478 \cellcolor[rgb]{0.9996, 0.9881, 0.7290} \\
Coverity & 2024.6.1 & 31 & 86 & 62 & 119 & 33\% & 21\% & 39\% & 425 \cellcolor[rgb]{0.9975, 0.9213, 0.6171} \\
Cppcheck & 2.13.0 & 19 & 100 & 9 & 131 & 68\% & 13\% & 46\% & 406 \cellcolor[rgb]{0.9967, 0.8975, 0.5771} \\
Clang Analyzer & 18.1.3 & 13 & 99 & 2 & 137 & 87\% & 9\% & 45\% & 381 \cellcolor[rgb]{0.9956, 0.8554, 0.5257} \\
GitLab SAST & 15.2.1 & 36 & 67 & 240 & 120 & 13\% & 23\% & 22\% & 374 \cellcolor[rgb]{0.9953, 0.8400, 0.5128} \\
Splint & 3.1.2 & 23 & 36 & 1027 & 127 & 2\% & 15\% & 5\% & -598 \cellcolor[rgb]{0.6471, 0.0000, 0.1490} \\
CodeThreat & - & 24 & 2 & 1101 & 126 & 2\% & 16\% & 2\% & -692 \cellcolor[rgb]{0.6471, 0.0000, 0.1490} \\
\hline \hline
GPT-o3 Mini & - & 126 & 60 & 73 & 36 & 63\% & 78\% & 63\% & 977 \cellcolor[rgb]{0.0000, 0.4078, 0.2157} \\
Chat GPT-o1 & - & 128 & 56 & 90 & 35 & 59\% & 78\% & 60\% & 962 \cellcolor[rgb]{0.0120, 0.4300, 0.2272} \\
DeepSeek R1 & - & 148 & 41 & 166 & 17 & 47\% & 90\% & 51\% & 956 \cellcolor[rgb]{0.0200, 0.4448, 0.2349} \\
GPT-4o & - & 136 & 45 & 116 & 43 & 54\% & 76\% & 53\% & 954 \cellcolor[rgb]{0.0240, 0.4521, 0.2388} \\
GPT-4o Mini & - & 134 & 27 & 263 & 43 & 34\% & 76\% & 34\% & 761 \cellcolor[rgb]{0.4492, 0.7627, 0.3936} \\
QWEN 2.5CI (32B) & - & 114 & 31 & 224 & 49 & 34\% & 70\% & 35\% & 708 \cellcolor[rgb]{0.5870, 0.8230, 0.4087} \\
Falcon 3 (7B) & - & 30 & 76 & 76 & 124 & 28\% & 20\% & 35\% & 521 \cellcolor[rgb]{0.9503, 0.9791, 0.6810} \\
Mistral Ins. (7B) & - & 63 & 23 & 215 & 91 & 23\% & 41\% & 22\% & 446 \cellcolor[rgb]{0.9982, 0.9452, 0.6571} \\
Gemma 2 (9B) & - & 63 & 42 & 258 & 95 & 20\% & 40\% & 23\% & 436 \cellcolor[rgb]{0.9979, 0.9356, 0.6411} \\
LLAMA 3.1 (8B) & - & 83 & 22 & 337 & 80 & 20\% & 51\% & 20\% & 417 \cellcolor[rgb]{0.9972, 0.9118, 0.6011} \\
\bottomrule
\end{tabular}
\smallskip

Legend: \textbf{TP} = True Positive; \textbf{TN} = True Negative; \textbf{FP} = False Positive; \textbf{FN} = False Negative; \textbf{P} = Precision; \textbf{R} = Recall; \textbf{A} = Accuracy;
\\
\textit{* Tools available exclusively through online APIs, without specified version numbers, were all evaluated in February 2025.}

\label{tab:results}
\end{table*}

The CASTLE Score is designed to provide a balanced assessment of a tool’s effectiveness by considering both true and false positives and the severity of vulnerabilities. Consequently, not finding a high-severity vulnerability leads to larger penalties than less impactful ones. A negative CASTLE Score could indicate that the volume of false positives generated by a tool imposes a significant triage burden on developers, outweighing its potential benefits. Overall, both the benchmark dataset and the introduced evaluation metric helped highlight various static analyzers' strengths and weaknesses.

\subsection{Tool evaluation on CASTLE Benchmark}

\begin{figure*}[htbp]
\centering
\includegraphics[width=1\textwidth]{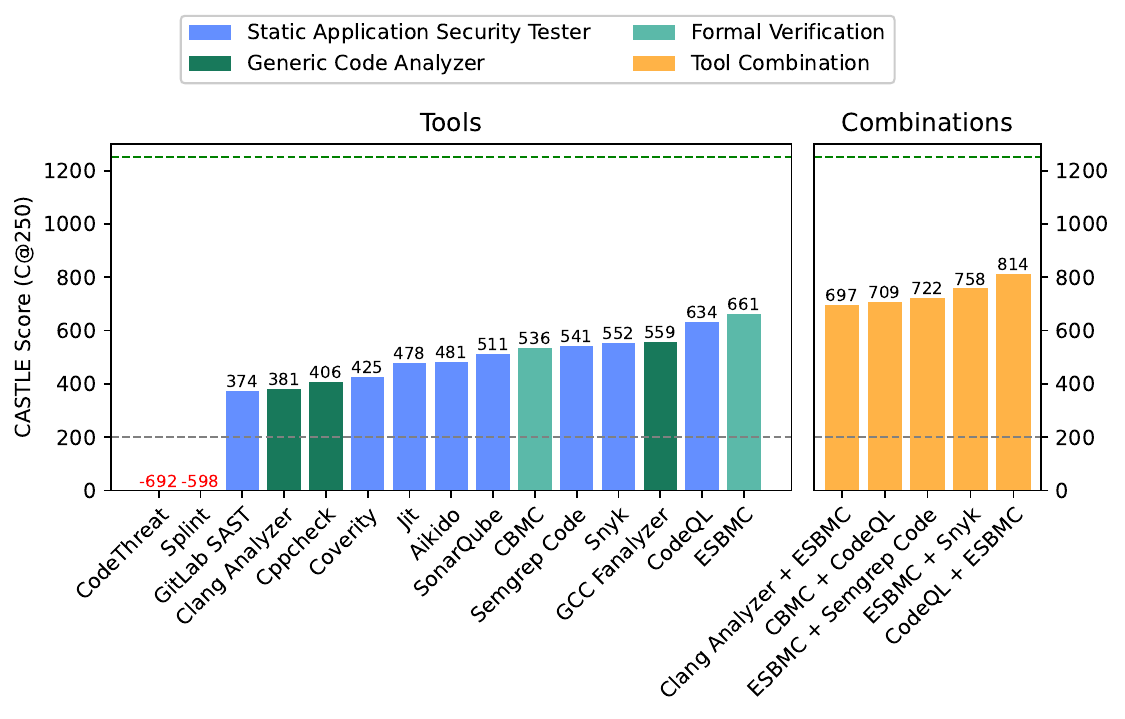}
\caption{CASTLE Scores for tools tested on $250$ C programs, including the top five tool combinations. Tools reporting no issues score $200$ points. The theoretical maximum of a perfect score is $1250$ points.}
\label{fig:castle-scores}
\end{figure*}

Figure~\ref{fig:castle-scores} shows the results of only the static analyzer tools and their best combinations. 

\begin{figure*}[htbp] 
\centering
\includegraphics[width=0.8\textwidth]{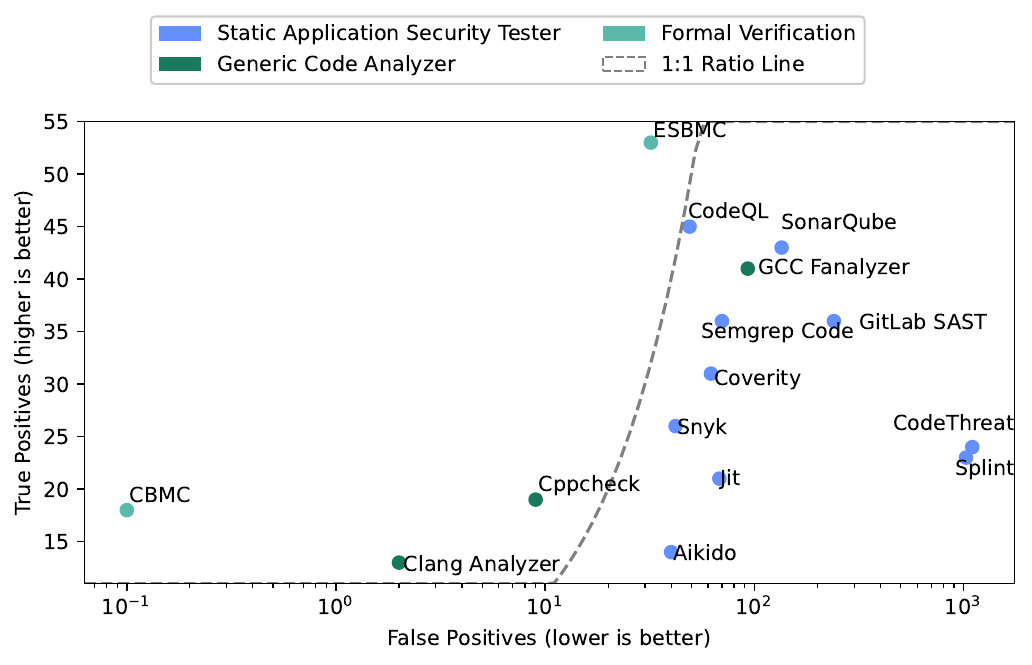}
\caption{True positive vs False positive rates of tools}
\label{fig:tp-fp-scatter}
\end{figure*}

The highest-performing individual tool in our analysis was ESBMC, a formal verification tool.
Formal Verification methods have the main disadvantage of being unable to detect non-formal issues, such as SQL Injection, Path traversal, or hard-coded credentials. However, they compensate for this with their low false positive rate. Theoretically, bounded model checkers cannot produce false positives, as they always provide a counterexample to their findings, except in cases where the tool itself has bugs or reports esoteric scenarios. CBMC achieved $0$ false positives, the only tool to do so (see Figure~\ref{fig:tp-fp-scatter}). ESBMC, however, had $32$ false positives in three categories, some of which we submitted as bug reports to the project \cite{esbmc_issue2312} \cite{esbmc_issue2236} \cite{esbmc_issue2235} \cite{esbmc_issue2231}. While this dataset with is short code samples allowed relaxed setting for ESBMC with a longer timeout; \texttt{--overflow --no-unwinding-assertions --memory-leak-check --timeout 60 --multi-property  --show-stacktrace}, with larger codebases formal verification tools could peotentially struggle to finish the verification process in reasonalbe time, thereby limiting their thoroughness and reliability in giving accurate results. The best-performing SAST tool is CodeQL, which found 29\% of the weaknesses in the code (45/150), the highest of the average of 23\% among other SASTs. SonarQube also received 29\%, but it was dragged down by reporting 2.7 times as many false positives than CodeQL. Several tools, including Clang Analyzer and CBMC, displayed extremely high precision (87\% and 100\% respectively) but struggled with low recall (9\% and 12\%). This trade-off implies they excel at correctly labeling the few issues they detect, yet they fail to identify a substantial portion of vulnerabilities. Conversely, ESBMC’s more balanced approach (62\% precision, 35\% recall) often provides a more reliable day-to-day detection rate for developers. Splint and CodeThreat generated exceptionally high false positives (1,027 and 1,101, respectively). Their negative CASTLE Scores (-598 and -692) illustrate how overwhelmingly false alerts can erode a tool’s practical utility. Although both tools still produced a modest number of true positives, the excessive manual triage effort likely outweighs any marginal benefits for most real-world projects.

Another advantage of the CASTLE score over traditional metrics is that it provides a comparison between using tool combinations. If a pair of tools has a high overlap in the CWEs they can detect, the CASTLE Score of their combination will yield a lower result than the individual tools because of the oversized impact of increasing the rate of false positives. When looking at combination scores, the biggest increase happens when ESBMC and CodeQL is combined, yielding 814 points. This is a 153 point increase over the higher performing ESBMC's base score of 661, and a 23\% increase in the effectiveness of using both tools instead of just ESBMC, with a 28\% increase above just using CodeQL. This shows that selecting tool combinations with different strengths significantly boosts the efficacy of the static analysis process.

\subsection{LLM Evaluation on CASTLE Benchmark}
\label{sec:llm_evaluation}

On the CASTLE dataset, LLMs exhibited notably strong performance. In particular, GPT-o3-Mini achieved the highest overall score of 977 points, correctly identifying 126 out of the 150 known vulnerabilities. When examining the true positives across different LLM variants, we observed that GPT-4o and GPT-4o-Mini generated more accurate detections than GPT-o1 or GPT-o3-Mini for true positives. However, the \emph{reasoning-oriented} models consistently produced fewer false positives, suggesting that their internal steps for ``validating'' potential vulnerabilities lead to more precise outcomes.

Our findings indicate that modern LLMs can pinpoint vulnerabilities in short, self-contained C programs. We conjecture that their neural architectures confer an inherent advantage in pattern recognition, while more advanced ``reasoning'' models appear better at avoiding out false detections. As a result, LLM-based approaches rival---and often surpass---several classical static analysis tools in detecting common software flaws within compact code segments. However, the next section highlights several limitations and issues for LLMs.

\section{Limitations}
\label{sec:limitations}

\textit{Microbenchmark Scope:}
A fundamental concern with any microbenchmark-based study is its limited scope. Although the CASTLE dataset covers 25 distinct CWEs, each test typically focuses on a single vulnerability in an isolated context. Real-world software often exhibits multi-faceted security flaws spanning tens or hundreds of files. Consequently, tools optimized for detecting specific vulnerabilities may perform artificially well on microbenchmarks while missing complex, cross-file weaknesses that only arise in large-scale applications. Regardless, tools did not perform well on even this small test, indicating that their high false positive rates would be a problem for longer contexts. 

\textit{Lack of Large Code Samples:}
Preliminary testing with a synthetic 400+ line C program created by merging multiple non-vulnerable test cases, revealed that LLMs tend to report false positives when dealing with larger codebases. Similarly, when one hidden vulnerability was introduced into this combined file, most LLMs failed to detect it reliably, suggesting that these models’ effectiveness may taper off with increasing code length. Formal verification approaches also suffer from scalability issues, such as state explosion, and may require lowered bounds that reduce their thoroughness. By contrast, classical static application security testers (SAST) can handle extensive projects more efficiently, yet their propensity for false positives undercuts overall usefulness in large-scale deployments.


\textit{Potential Overfitting:}
Because CASTLE test contents are fixed, tool vendors could theoretically fine-tune their analyzers to excel on known benchmarks, inflating reported accuracy while not generalizing to unseen software. Although this consideration does not impact the integrity of our current study, it underscores the importance of periodically refreshing the dataset or incorporating dynamic test-generation approaches for the future. Furthermore, while repeated evaluations of the same code typically yield consistent results (with observed deviations below 3\%), the inherent stochasticity of model-based systems stands in contrast to the deterministic nature of many static analyzers.

\section{Conclusion}
\label{sec:conclusion}

\noindent In this study, we introduced the CASTLE benchmark, a curated collection of 250 compilable C micro-benchmarks covering 25 major CWEs. We proposed the CASTLE Score to evaluate diverse vulnerability detection tools, including static analyzers, formal verification methods, and large language models (LLMs). Our work aimed to address the following research questions:

\begin{itemize} \small
    \item \textbf{RQ1:} \textit{How do state-of-the-art static analysis tools, formal verification methods, and LLM-based approaches compare in effectively detecting vulnerabilities in C code?} \\
    \textbf{Answer:} 
    LLMs exhibit high effectiveness on compact code snippets, with GPT-o3-Mini scoring the highest (977 points) by identifying 126 out of 150 vulnerabilities. However, their performance declines on larger codebases, where false positives increase and hidden vulnerabilities often remain undetected. Static analyzers perform moderately but produce numerous false positives, creating substantial manual triage overhead. Formal verification tools yield minimal false positives within their supported classes (e.g., memory safety) but cannot detect certain higher-level vulnerabilities such as SQL injection, limiting their coverage.

    \item \textbf{RQ2:} \textit{Are combinations of tools more effective than using a single tool?} \\
    \textbf{Answer:}
    Tool combinations frequently outperform individual tools, particularly when they offset each other’s weaknesses. For instance, ESBMC (formal verification) combined with CodeQL achieved the highest two-tool CASTLE Score (814). Although overlapping detections can inflate false positives, well-chosen pairs leverage complementary detection strategies, enhancing overall reliability.

    \item \textbf{RQ3:} \textit{What metrics can reliably demonstrate these differences among various tools?} \\
    \textbf{Answer:}
    As shown in Table~\ref{tab:results}, neither precision, accuracy, or recall could have produced the same results and insights. The CASTLE Score integrates true positives, false positives, and CWE severity, providing a single, clear measure of tool performance. This setup enables transparent evaluation and straightforward comparisons across diverse methods, even for tool combinations.
\end{itemize}

\noindent \textbf{Implications and Future Work.}
Although micro-benchmarks efficiently reveal how tools behave on targeted vulnerabilities, they may not reflect the full complexity of production-scale systems. Preliminary experiments show that LLMs and formal verification tools both face significant scalability barriers when analyzing large codebases. Ultimately, the insights gained through CASTLE underscore the importance of selecting and combining tools to fit specific project requirements rather than relying on any single method for comprehensive security assurance.

\section{Acknowledgement}
\label{sec:acknowledgements}

This research is partially funded and supported by ZEISS Digital Innovation. Additionally, it receives partial support through the TKP2021-NVA Funding Scheme under Project TKP2021-NVA-29, as well as the Research Council of Norway under Project No. 312122, “Raksha: 5G Security for Critical Communications.”

\bibliographystyle{IEEEtran}
\bibliography{main} 

\end{document}